\documentclass[onefignum,onetabnum]{siamart171218}
\usepackage[toc,page]{appendix}
\usepackage[shortlabels]{enumitem}
\usepackage{color}

\newcommand{\rank}{\mathrm{rank}}

\usepackage{lipsum}
\usepackage{amsfonts}
\usepackage{epstopdf}
\ifpdf
  \DeclareGraphicsExtensions{.eps,.pdf,.png,.jpg}
\else
  \DeclareGraphicsExtensions{.eps}
\fi

\usepackage{amsmath}
\usepackage{amssymb}
\usepackage{float}
\usepackage{graphicx}
\usepackage{color}
\usepackage{multirow}
\usepackage{tabularx}
\usepackage{textcomp}
\usepackage{caption}
\usepackage{algpseudocode}

\newcommand{\R}{\mathbb{R}}

\newcommand{\ttext}[1]{\text{\tiny{#1}}}

\newcommand{\role}[1]{\texttt{#1}}

\DeclareMathOperator*{\argmin}{arg\,min}

\newenvironment{compactitemize}{
    \begin{itemize}\vspace{-\topsep}
  }{
    \vspace{-\topsep}\end{itemize}
  }

\newsiamremark{remark}{Remark}
\newsiamremark{hypothesis}{Hypothesis}
\crefname{hypothesis}{Hypothesis}{Hypotheses}
\newsiamthm{claim}{Claim}

\headers{Knowledge graph recommendations}{M. Beguerisse, D. Korkinof, and T. Hoffmann}

\title{Thematic recommendations on knowledge graphs using multilayer 
    networks\thanks{Submitted to the editors \today.}}

\author{Mariano Beguerisse-D\'iaz\footnotemark[2]\ \footnotemark[3]
\and Dimitrios Korkinof\footnotemark[2]\ \footnotemark[3]
\and Till Hoffmann}

\usepackage{amsopn}

\makeatletter
\newcommand*{\addFileDependency}[1]{
  \typeout{(#1)}
  \@addtofilelist{#1}
  \IfFileExists{#1}{}{\typeout{No file #1.}}
}
\makeatother

\ifpdf
\hypersetup{
  pdftitle={Thematic recommendations on knowledge graphs using multilayer networks},
  pdfauthor={M. Beguerisse-Diaz, D. Korkinof, and T. Hoffmann}
}
\fi

\begin{document}

\maketitle

\renewcommand{\thefootnote}{\fnsymbol{footnote}}

\footnotetext[2]{Spotify Ltd, UK. (\email{marianob@spotify.com}, \email{dkorkinof@spotify.com})}
\footnotetext[3]{These authors contributed equally to this work}

\begin{abstract}

   We present a framework to generate and evaluate thematic recommendations based on multilayer network representations of knowledge graphs (KGs). In this representation, each layer encodes a different type of relationship in the KG, and directed interlayer couplings connect the same entity in different roles. The relative importance of different types of connections is captured by an intuitive salience matrix that can be estimated from data, tuned to incorporate domain knowledge, address different use cases, or respect business logic.

  We apply an adaptation of the personalised PageRank algorithm to multilayer models of KGs to generate item-item recommendations. These recommendations reflect the knowledge we hold about the content and are suitable for thematic and/or cold-start recommendation settings. Evaluating thematic recommendations from user data presents unique challenges that we address by developing a method to evaluate recommendations relying on user-item ratings, yet respecting their thematic nature. We also show that the salience matrix can be estimated from user data. We demonstrate the utility of our methods by significantly improving consumption metrics in an AB test where collaborative filtering delivered subpar performance. We also apply our approach to movie recommendation using publicly-available data to ensure the reproducibility of our results. We demonstrate that our approach outperforms existing thematic recommendation methods and is even competitive with collaborative filtering approaches.
\end{abstract}



\section{Introduction}
\label{sec:intro}

Knowledge graphs (KGs) encode semantic relationships between large collections of items~\cite{Ehrlinger2016, Hogan2020}. In recent years, there has been a surge of interest in KGs in both academic and industrial settings. Much of this interest is due to a shift
from information to knowledge retrieval, and initiatives like the Google knowledge vault~\cite{dong2014knowledge} or Amazon's product graph~\cite{dong2018challenges}. Typical applications of KGs include reasoning~\cite{Bellomarini2018,Chen2020}, search~\cite{Yang2016}, and recommendations~\cite{Yu2014}.

Recommendation systems fall into three broad categories: (i) collaborative filtering (CF) approaches~\cite{Shi2014}, formulated as a missing data problem to predict the likely interaction between a user and an item, (ii) reinforcement learning (RL)
approaches~\cite{Zheng2018,mcinerney2018explore} to continuously refine recommendations in an online setting, and (iii) knowledge-based approaches, which exploit the information we hold about items to generate recommendations~\cite{Shi2014,Covington2016}. Graph-based approaches often fall into the latter category, and knowledge is captured by relationships among items and other entities (e.g. tags or users)~\cite{Hogan2020}.  Some recommender systems fuse multiple approaches to further improve recommendations: for example, CF and KGs~\cite{monti2017geometric} or KGs with RL~\cite{xian2019reinforcement}.

Collaborative filtering systems predict which items users are most likely to interact with based on their past interactions with other items~\cite{Lue2012, Ricci2010}. The success of CF systems has made them a mainstay in research and practice of recommender systems. However, there are situations where user-item interactions alone are not enough to achieve the desired performance~\cite{Covington2016}, or sufficient interaction data may
not be available (e.g. a cold-start recommendation setting)~\cite{Gope2017}. Several recommendation systems thus combine collaborative filtering approaches with annotations and KGs to enrich their recommendations~\cite{Szomszor2007,Wang2010}.

A common approach to recommendation systems based on KGs is to generate node embeddings using graph networks or matrix factorisations~\cite{Ai2018, Dong2017, Rossi2020, Salha2019, Sun2018, Ying2018, Yu2014}. Although embedding methods can achieve impressive performance, their recommendations can be difficult to explain. This has motivated researchers to work on methods to understand and explain recommendations~\cite{Ai2018}. Furthermore, the geometry of embedding spaces can be complex~\cite{Chang2017}, which presents difficulties in tasks such as proximity search and multi-seeding, i.e. using multiple items to provide a recommendation context. The lack of interpretability furthermore poses challenges for assessing fairness and bias in complex machine learning systems~\cite{Cramer2018, Poursabzi-Sangdeh2018}. From a modelling perspective, one concern that arises in some embedding methods is that they may not fully exploit the richness of the structure of the KG~\cite{Rossi2020}, or that they do not adequately distinguish between connections of different type in the KG~\cite{Yu2014}. Some methods do use the graph structure to generate recommendations, e.g. using random walks and diffusion processes~\cite{Wang2010, Chaudhari2017, Nikolakopoulos2019, Eksombatchai2018}; however, these works do not often make a distinction between different types of connections, which can have an important effect on the quality of the recommendations.

In this work we propose a principled method to generate \emph{thematic} item-item recommendations using random walks on multilayer network representations of KGs. Broadly, we say recommendations are thematic if they rely on the intrinsic properties of the items, which can be encoded as connections and nodes in a KG. Our method enables thematic exploration of items in the KG with or without user data, making it applicable to cold-start recommendation
problems. In addition, our method allows us to exploit the full structure of the KG, accounting for the distinction between different types of connections.

This paper is structured as follows: in Sec.~\ref{sec:network_mods}, we show how to represent a KG as a multilayer network where each layer represents one connection type; nodes that represent the same entity in different layers are connected to each other by directed, weighted interlayer couplings. Then, in Sec.~\ref{sec:random} we construct the rate matrix of a random walk on the multilayer network, incorporating a set of parameters that encode the salience of each connection type. These saliences put different connection types on an equal footing, facilitating direct comparison between connections. In Sec.~\ref{sec:recs}, we use an adaptation of personalised PageRank to produce recommendations using arbitrary sets of nodes as seeds.  We introduce a method to evaluate thematic recommendations based on user data in Sec.~\ref{sec:eval}; this evaluation differs from traditional CF evaluation approaches because it assesses recommendations generated using the connections in the KG only. In Sec.~\ref{sec:applications}, we showcase two applications of our method: in Sec.~\ref{sec:ab-test}, we show the results of an AB test for thematic music recommendations on a cohort of 100k users, and in Sec.~\ref{sec:tmdb_application}, we apply our method to a KG of the film industry, which we evaluate offline using the method from Sec.~\ref{sec:eval}. We show that we can learn the value of the salience parameters by optimising this evaluation score. These saliences can be interpreted as the relative importance of the different types of connections in the KG, and can be used to explain the output of the system.  We compare our approach with alternative methods and show that our recommendations can outperform other thematic recommendation systems and be competitive with CF methods. Finally, in Sec.~\ref{sec:discussion}, we summarise our methods and results, and indicate potentially interesting avenues for future research and applications.

\section{Network models of knowledge graphs}
\label{sec:network_mods}

\begin{figure}[tp]
  \centering
  \includegraphics[width=\textwidth]{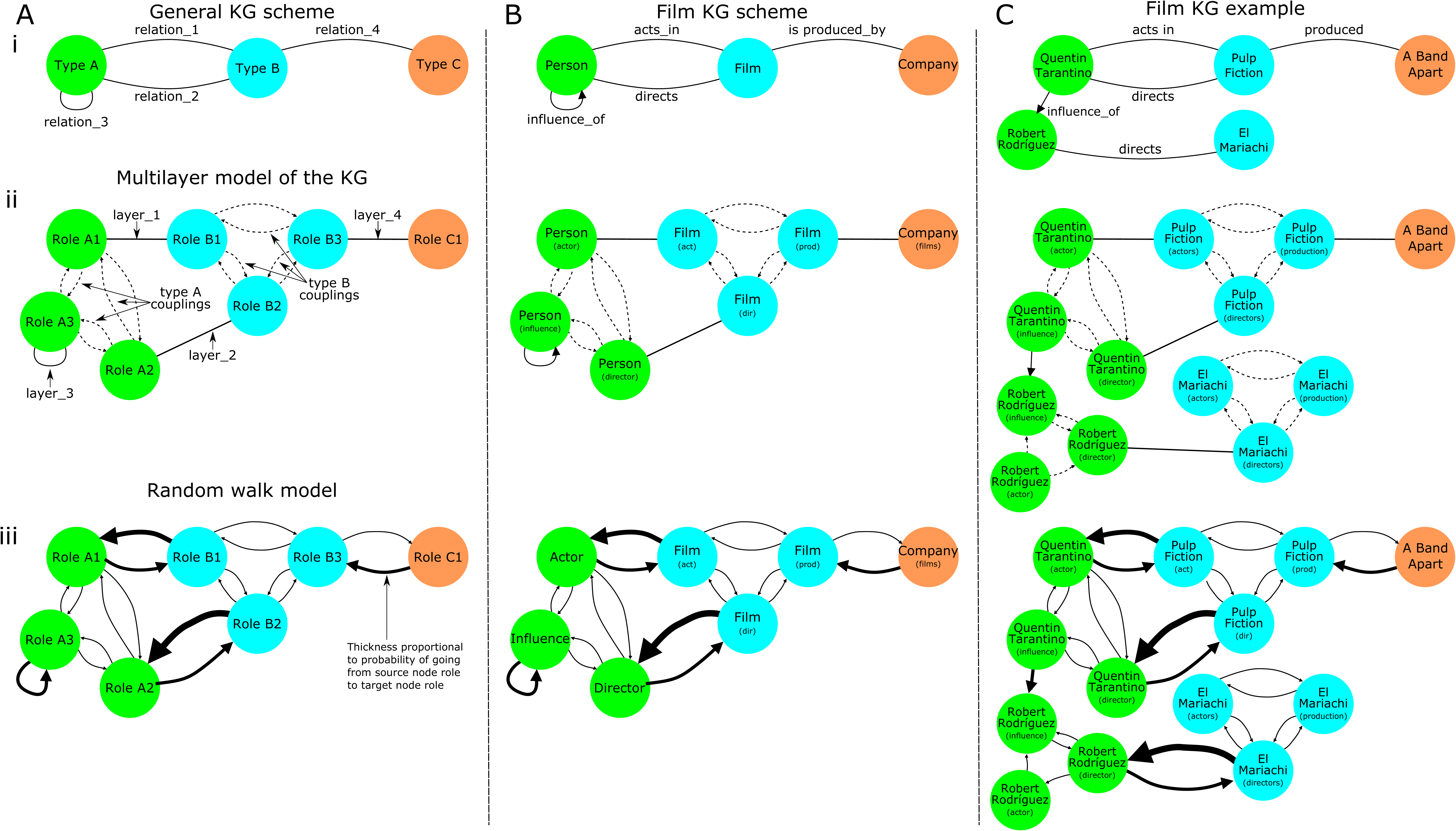}
  \caption{{\bf A}: (i) A general schematic of a KG with three types
    of entities and four types of relations. (ii) Multilayer network
    representation of the KG. Each layer corresponds to a type of
    relation in the KG (solid lines). Unipartite layers model
    interactions among nodes representing the same role, and bipartite
    layers model interactions between two different roles. Nodes that
    represent the same entity in different layers are coupled to each other (dashed
    lines). (iii) Illustration of a random walk model on the
    multilayer representation of the KG. In this diagram, the width of
    each directed connection indicates the proclivity of a random
    walker to go from a node in the source role to another in the
    target role. {\bf B}: a schematic of a film KG with three types of
    entities (person, film, company), and four types of relations
    (acting, directing, influencing and producing). In the multilayer
    representation of the KG, there are three roles for each person
    entity (director, actor, influence), three roles for films (one
    for each of its interactions with actors, directors and
    companies), and one role for companies. {\bf C}: a specific example of a
    film KG, its multilayer representation, and random walk model.}
  \label{fig:toy_kg}
\end{figure}

Knowledge graphs (KGs) are typically \emph{multigraphs}: entities, represented by nodes, are connected by different types of relationships~\cite{Diestel2005}. Figure~\ref{fig:toy_kg} shows an example of a schematic of a KG about the film industry. This KG comprises three types of entities: person, film, and company. Companies produce films; a person can contribute to a movie by either directing it or acting in it, and influence other people's work. These types of relations encode distinct information and are thus neither interchangeable nor directly comparable. For example, na\"ively evaluating the centrality of nodes in a KG by treating all relationships in the same manner would discard important information, and may lead to suboptimal results. A recommendation algorithm using KGs should account for the different types of relationships.

To make the best use of all the information encoded by the
relationships in the KG, yet develop a tractable thematic
recommendation system, we adopt a multilayer network
approach~\cite{Kivela2014}: each relationship type is represented by one
layer, and the same entity in the KG can appear in different
layers, fulfilling different roles. Henceforth, we will use the term
\emph{role} to refer to how an entity in the multilayer network
relates to others (e.g. a person acting in or directing a movie) and reserve the
term entity \emph{type} to refer to what a node is (e.g. a person,
company or movie).

The knowledge graph in Fig.~\ref{fig:toy_kg} encodes information about a
set $\mathcal{P}$ of person entities, $\mathcal{C}$ of company
entities, and $\mathcal{F}$ of film entities with cardinality $P$,
$C$, and $F$, respectively. Person and company entities play one of
the following roles:
\begin{compactitemize}
  \item people acting in movies are represented by \role{actor} roles (\role{p-act}),
  \item people directing movies are represented by \role{director} roles (\role{p-dir}),
  \item people influencing others are represented by \role{influencer} roles (\role{p-inf}),
  \item and companies producing movies are represented by \role{production company} roles (\role{c-prod}).
\end{compactitemize}
The passive involvement of movies is represented by the following
complementary roles:
\begin{compactitemize}
  \item \role{film-acting} for films connected to actors (\role{f-act}),
  \item \role{film-directing} for films connected to directors (\role{f-dir}),
  \item and \role{film-producing} for films connected to production companies (\role{f-prod}).
\end{compactitemize}
Nodes in the multilayer network that represent the same entity, such as
Quentin Tarantino being both the director of and an actor in the 1994
film Pulp Fiction, are connected to one another by interlayer
couplings. Figure~\ref{fig:toy_kg}B provides an illustration of the entities
and relationships outlined above.

More formally, we consider a multilayer network of the film KG in
Fig.~\ref{fig:toy_kg} with four layers, one for each relationship type,
and $N = 3 F + 3 P + C$ nodes across all layers: one node for each
entity in all roles in which it \emph{could} be active (e.g. an
actor who does not direct still has a director node).  The
supra-adjacency matrix $A\in \R^{N\times N}$, i.e. a flattened
representation of the multilayer network~\cite{Kivela2014}, is defined by
\begin{equation}
  A = B + \Lambda,
  \label{eq:supra-adjacency}
\end{equation}
where $B$ is a block matrix representing the relationships of the
knowledge graph. Each block $B^{(l)}$ in $B$ is the adjacency matrix of the
layer encoding relationship type $l$ (i.e. the \emph{intralayer}
connections in a multilayer network). The size of $B^{(l)}$ depends on
the entity types that participate in $l$. For example, the \role{directs} block represents a bipartite graph and has $P$ rows and $F$ columns. The weight $B_{ij}$
of the connection between two nodes $i,\,j\in \{1, \dots, N \}$
encodes the a priori importance of the relationship, such as the
fraction of screen time an actor has in a movie. If the nodes are not
connected, then $B_{ij} = 0$.

The block matrix $\Lambda$ captures the directed, weighted
\emph{interlayer} couplings between all nodes that represent the same
entity (illustrated by dashed lines in Fig.~\ref{fig:toy_kg}). The
directed interlayer coupling from $j$ to $i$ exists if the two nodes
represent the same entity in different roles, and the corresponding weight is equal to
the weighted degree of node $i$ within its layer. Consequently, nodes
that are not active in a given role have no incoming interlayer
connections, but do have outgoing ones. For example, a connection from Tarantino's 
actor node to
Tarantino's director node exists because they represent the same
person \emph{and} he has directed movies. In contrast, Robert Rodr\'{i}guez
has no acting credits (in this example), thus his actor node has no
incoming interlayer connections, as shown in Fig.~\ref{fig:toy_kg}C.
Each block of $\Lambda$ is a square matrix $\Lambda^{(r)}$ whose main 
diagonal contains the weighted degree of all nodes with role $r$, and has 
zeroes everywhere else. The size of each block depends on the entity type 
that assumes the role. For example, $\Lambda^{(\mathrm{p-dir})}$ has size 
$P \times P$, and $\Lambda^{(\mathrm{p-dir})}_{ij}=
k_i^{(\mathrm{p-dir})}\delta_{ij}$ for $i,j \in \{1, \dots, P \}$ where
$k_i^{(\mathrm{p-dir})}$ is the weighted degree of person $i$ as
director and $\delta_{ij}$ is the Kronecker delta.

Finally, the block structure of the supra-adjacency matrix of the
multilayer network is
\begin{equation}
\small
 A =\quad
 \bordermatrix{
&\mbox{\tiny f-act} &\mbox{\tiny f-dir} &\mbox{\tiny f-prod} & \mbox{\tiny p-act} &\mbox{\tiny p-dir} &\mbox{\tiny p-infl} &\mbox{\tiny c-prod} \cr
\mbox{\tiny f-act} &   & \Lambda^{(\mathrm{f-act})}  & \Lambda^{(\mathrm{f-act})}  & {B^{(\mathrm{act})}}^\intercal &  &  & \cr
\mbox{\tiny f-dir} & \Lambda^{(\mathrm{f-dir})} &  & \Lambda^{(\mathrm{f-dir})}  &  & {B^{(\mathrm{dir})}}^\intercal  & & \cr
\mbox{\tiny f-prod} & \Lambda^{(\mathrm{f-prod})}  & \Lambda^{(\mathrm{f-prod})} & &  & &  & {B^{(\mathrm{prod})}}^\intercal \cr
\mbox{\tiny p-act} & B^{(\mathrm{act})} &  &  &  & \Lambda^{(\mathrm{p-act})}  &  \Lambda^{(\mathrm{p-act})}  & \cr
\mbox{\tiny p-dir} &  & B^{(\mathrm{dir})}  &  & \Lambda^{(\mathrm{p-dir})} &   &   \Lambda^{(\mathrm{p-dir})}  & \cr
\mbox{\tiny p-inf} &  &   &  & \Lambda^{(\mathrm{p-inf})} &   \Lambda^{(\mathrm{p-inf})} & B^{(\mathrm{infl})}  & \cr
 \mbox{\tiny c-prod} &  &  &  B^{(\mathrm{prod})} & & & & }.
  \label{eq:supra-adjacency-example}
\end{equation}
This matrix encodes the KG in Fig.~\ref{fig:toy_kg} \emph{without loss of
any information}. For example, if Quentin Tarantino and Robert
Rodr\'iguez are $i,\,k\in\mathcal{P}$ respectively, and Pulp Fiction
is $j\in\mathcal{F}$, the acting connection between Tarantino and Pulp
Fiction in Eq.~\eqref{eq:supra-adjacency-example} 
is found in $A_{j,\,3 F + i}$ and $A_{3 F + i,\,j}$ (via
$B^{\mathrm{(act)}}_{ij}$); the directing connection is in $A_{F + j,\,3 F
  + P + i}$ and $A_{3 F + P + i,\,F + j}$ (via $B^{\mathrm{(dir)}}$);
Tarantino's influence on Rodr\'iguez is only in entry $A_{3F + 2P +
  k,\, 3F + 2P + i}$ (via $B^{\mathrm{(infl)_{ik}}}$) because influence is
a directed layer. Finally, the fact that Tarantino the actor, the
director, and the influence are the same person is encoded in six
elements of $A$ (via $\Lambda^{\mathrm{(p-dir)_{ii}}}$,
$\Lambda^{\mathrm{(p-act)_{ii}}}$ and $\Lambda^{\mathrm{(p-inf)_{ii}}}$). Note
that an entity that is not active in a specific role still has a node
for it; this node has no incoming connections, and the only outgoing
connections are interlayer couplings to itself in roles in which the
entity is active. An assumption we make throughout this work is that
all entities are active in at least one layer (i.e. there are no
isolated entities in the KG).

We can reduce the size of the multilayer model by merging
roles. For example, if we merge all movie roles into a single one, the supra adjacency of the multilayer network would be
\begin{equation}
\small
 A' = \quad
\bordermatrix{
     & \mbox{\tiny film} & \mbox{\tiny p-act} & \mbox{\tiny p-dir} & \mbox{\tiny p-infl} & \mbox{\tiny c-prod}\cr
     \mbox{\tiny film} & & {B^{(\mathrm{act})}}^\intercal & {B^{(\mathrm{dir})}}^\intercal & & {B^{(\mathrm{prod})}}^\intercal \cr
     \mbox{\tiny p-act} & {B^{(\mathrm{act})}}  & & \Lambda^{(\mathrm{p-act})}  &  \Lambda^{(\mathrm{p-act})}  & \cr
     \mbox{\tiny p-dir} & {B^{(\mathrm{dir})}} &\Lambda^{(\mathrm{p-dir})} &   &   \Lambda^{(\mathrm{p-dir})}  & \cr
     \mbox{\tiny p-infl}&& \Lambda^{(\mathrm{p-inf})} &   \Lambda^{(\mathrm{p-inf})} & B^{(\mathrm{infl})}  & \cr
     \mbox{\tiny c-prod}&{B^{(\mathrm{prod})}} & & & &
},
  \label{eq:reduced-supra-adjacency-example}
\end{equation}
with $F+3P+C$ nodes. In $A'$, all the interactions with movies are in the first 
block row/column, and that there are no interlayer couplings $\Lambda$ 
between movie nodes. 
The most aggressive merging strategy would create
a single role for each entity (i.e. one node per person, one per film, and so on), and would have $F + P + C$ nodes (compared with $3F + 3P + C$ nodes in the full model):
\begin{equation}
\small
 A'' = \quad
\bordermatrix{
      & \mbox{\tiny film} & \mbox{\tiny person} & \mbox{\tiny company}\cr
     \mbox{\tiny film} & & {B^{(\mathrm{act})}}^\intercal + {B^{(\mathrm{dir})}}^\intercal & {B^{(\mathrm{prod})}}^\intercal \cr
     \mbox{\tiny person} & B^{(\mathrm{act})} + B^{(\mathrm{dir})} & B^{(\mathrm{infl})}\cr
     \mbox{\tiny company} & B^{(\mathrm{prod})} \cr
}.
  \label{eq:reduced-supra-adjacency-example-VERSION2}
\end{equation}
Merging roles can have computational and modelling benefits in certain applications. However, because connections are not generally comparable (e.g. acting is not the same as directing), roles need to be merged with care. Furthermore, it may not be possible to recover the full KG from a reduced mutilayer model; i.e. some information may be lost.

This modelling approach defines a family of multilayer networks in which layers and roles are determined by the subset of the relations from the KG that we choose to include in a particular application. The choice of which layers to include must be carefully considered by the modeller, taking into account quality of the data and the application at hand. Adding unnecessary layers may result in larger models that are more difficult to analyse and perform poorly. An advantage of this approach is that multilayer representations of KGs benefit from the wealth of theory and methods from the multilayer network literature (see Ref.~\cite{Kivela2014} for a review article).

\section{Random walks on multilayer models of knowledge graphs}
\label{sec:random}

The study of dynamics on networks has proven useful for information retrieval tasks~\cite{Gleich2015, Higham2003, Page1998}, and random walks on networks~\cite{Masuda2017} provide a convenient and effective framework to explore graph structures for recommendations~\cite{Dong2017, Nikolakopoulos2019, Ying2018}. Here we use the properties of random walks on multilayer models of KGs to generate \emph{thematic} recommendations, as defined in Sec.~\ref{sec:intro}. We interpret the ``thematic relatedness'' of two nodes as how easily reachable they are in the KG. 

A random walker on the multilayer network can transition between nodes
by following any of the connections encoded in the supra-adjacency
matrix $A$, as illustrated in row~(iii) of Fig.~\ref{fig:toy_kg}. For example, a
walker on a \role{f-act} node may step to either an actor featured in
the film, or another role of the same film, such as \role{f-prod} or
\role{f-dir}. Similarly, a walker on an \role{actor} node can transition to
one of the films connected to the actor or to the same person in
another role, such as director or influencer.

Whilst the couplings in the supra-adjacency matrix $A$ in
Eq.~\eqref{eq:supra-adjacency} encode the a priori importance of each
relationship, connections of different types are not necessarily
directly comparable (e.g. edge weights may have different meaning, units or
scale). It would thus not be appropriate to obtain the transition matrix of a random walk by simply column-normalising this matrix.
The relative importance of different relationship types
is often unknown a priori and may differ by use case. For example, we may
value a directing connection more than an acting one or vice versa
depending on the application.

To bring all relationships on the same footing 
and make them directly comparable (i.e. same units and
scale), we introduce a non-negative
salience matrix $\alpha\in\R^{N\times N}$ with the same block structure as $A$:
\begin{equation}
\arraycolsep=1pt
\small
  \alpha \hspace{-1pt} = \hspace{-2pt} \left( \hspace{-3pt}
\begin{array}{lllllll}
 & \alpha^\ttext{(f-act,f-dir)}  & \alpha^\ttext{(f-act,f-prod)}   & \alpha^\ttext{(f-act,p-act)} &  &  & \\
 \alpha^\ttext{(f-dir,f-act)} &  & \alpha^\ttext{(f-dir,f-prod)}   &  & \alpha^\ttext{(f-dir,p-dir)}  & & \\
 \alpha^\ttext{(f-prod,f-act)}  &  \alpha^\ttext{(f-prod,f-dir)} & &  & &  & \alpha^\ttext{(f-prod,c-prod)} \\
\alpha^\ttext{(p-act,f-act)} &  &  &  & \alpha^\ttext{(p-act,p-dir)}  &  \alpha^\ttext{(p-act,p-inf)}  & \\
& \alpha^\ttext{(p-dir,f-dir)} &  &  \alpha^\ttext{(p-dir,-p-act)}  &   &   \alpha^\ttext{(p-dir,p-inf)}  & \\
 &  &  & \alpha^\ttext{(p-inf,p-act)}   & \alpha^\ttext{(p-inf,p-dir)} &  \alpha^\ttext{(p-inf,p-inf)} & \\
 &  &  \alpha^\ttext{(c-prod,f-prod)} & & &  &
\end{array}
 \hspace{-4pt}
\right).
 \label{eq:salience-matrix}
\end{equation}
The entries of $\alpha$ encode the relative salience of the
connections and couplings originating from each node. For example,
consider the role \role{p-act}: the entries of
$\alpha^\ttext{(f-act,p-act)}$ encode the salience of connections from
\role{p-act} nodes to \role{f-act} nodes. We can compare these values
to the other blocks in the same column such as $\alpha^\ttext{(p-dir,
  p-act)}$, which encodes the salience of the coupling from actors to
themselves as directors.

The matrix $\alpha$ is not necessarily symmetric because transitions
in one direction may be more important than the opposite. We use a
shared salience value within each block,
i.e. ${\alpha^{(r_1,r_2)}}_{ij}=c^{(r_1,r_2)} \in \R^+$ for all $i$
and $j$, and roles $r_1$ and $r_2$, although it is possible to encode 
saliences for each node, or
even individual connections in the most general case. Note that the
connections in the KG may be dimensional, so the units of the
corresponding $\alpha^{(r_1,r_2)}$ matrix should be the inverse of the
connections'. For example, if the weight of connections from
\role{actor} to \role{film-acting} nodes encodes minutes on screen,
the units of the corresponding entries of
$\alpha^\ttext{(f-acting,act)}$ are $\mathrm{min}^{-1}$.

The first step to construct the transition matrix of a random walk is
to combine both the salience of different relationship types and the
knowledge encoded in the graph. We define the matrix
\begin{equation}
  \widetilde{T} = A\circ\alpha,
  \label{eq:supra-salience-elementwise}
\end{equation}
where $\circ$ denotes the element-wise product.  Because we combine the a
priori connection strength and salience, the entries of
$\widetilde{T}$ are non-dimensional and directly comparable within
each column, which means that we can now normalise its columns to construct the
rate matrix of a random walk.
The probability that a walker transitions from node $j$ to node $i$ is
\begin{equation}
  T_{ij} = \frac{\widetilde{T}_{ij}}{\sum_{i=1}^N
    \widetilde{T}_{ij}},
    \label{eq:transition-normalisation}
\end{equation}
which defines a matrix $T$ equal to $\widetilde{T}$ after column-normalisation to guarantee that the density of walkers is conserved;
i.e. $T$ is column-stochastic.  These probabilities depend only on the
KG data and the salience matrix $\alpha$ in
Eq.~\ref{eq:salience-matrix}. Therefore, we can exercise some control on 
the dynamics
of a random walk by tuning $\alpha$; this can be useful in contexts
where different aspects of the data become more or less important
(e.g. emphasising directing connections over acting in one context,
and vice versa in another). See Refs.~\cite{DeDomenico2014,
  Taylor2020} for similar ways to parametrise random walks on
multilayer networks.

To rank nodes in the multilayer model of the KG, we adopt the widely-used
personalised PageRank model of Ref.~\cite{Page1998}: a walker follows a
link with probability $1-\rho\in [0,1]$ and ``teleports'' to another
node with probability $\rho$. The probability to transition to a
particular node via teleportation is encoded in the $N\times 1$ vector
$v$, where $v_i\geq 0$ is the probability that node $i$ receives a
teleported walker, and $\sum_{i}v_i=1$. Let $x_i(t)$ denote the
probability that a walker is present on node $i$ at time $t$; this
vector evolves according to the transition rule
\begin{equation}
  x(t + 1) = (1-\rho) T x(t) + \rho v.\label{eq:teleportation-evolution}
\end{equation}
The PageRank vector is the steady state of the distribution of walkers
$\pi$, which occurs when $x(t+1)=x(t)=\pi(v)$. The value of each
element $\pi_i(v)$ is the fraction of time that a random walker spends
on node $i$ in an infinitely long random walk, given a teleportation
vector $v$.  Substituting $\pi(v)$ into
Eq.~\ref{eq:teleportation-evolution}, we obtain the linear
system~\cite{DelCorso2005}
\begin{equation}
  (I - (1-\rho) T)\pi(v) = \rho v, \label{eq:ppr-linear}
\end{equation}
where $I$ is the identity matrix.

Intuitively, the teleporation vector $v$ encodes the seeds of a walker
on the multilayer network, and $\rho$ captures the curiosity of the
walker, i.e., the likelihood it will explore the network. In the limit
$\rho \to 0$, we recover eigenvector centrality, and the steady state
state is trivially equal to the teleportation vector when $\rho =
1$. While the steady state solution can in principle be obtained by
iterating Eq.~\ref{eq:teleportation-evolution} until convergence, the
formulation in Eq.~\ref{eq:ppr-linear} is preferable because we can
use state of the art numerical techniques to solve the linear
system~\cite{DelCorso2005}. See Ref.~\cite{Gleich2015} for a review on
PageRank.


\section{Thematic recommendations on multilayer networks}
\label{sec:recs}

The fundamental assumption underpinning this work is that the
connections in the KG are relevant for recommendations. Given an item or set of items
of interest (i.e. the seeds), we can retrieve other items in the KG that
are related to them via the graph's connections.  We can encode the
recommendation context in the teleporation vector $v$ by giving seed
nodes a higher probability of receiving teleported walkers. For
example, the teleportation vector appropriate for a user who has
expressed interest in both Quentin Tarantino and Samuel L. Jackson
will assign substantial weight to their corresponding
nodes. Consequently, the steady state distribution of the random walk
$\pi(v)$ in Eq.~\eqref{eq:teleportation-evolution} will be biased in favour
of nodes that are easily reachable from those seeds. The value of $\pi_i(v)$
can be interpreted as a contextual score of importance for node $i$
given the seed vector $v$. Thus, we can obtain a ranked list of
recommendations as
$$ z'(v) = \mathrm{argsort}(-\pi(v)). $$
For our hypothetical Tarantino-Jackson aficionado, Pulp Fiction is
likely to have a high probability of being visited given the seed
vector, and it would appear near the top the ranked list of nodes.

More generally, the teleportation vector $v$ depends on the
recommendation context, and it may be determined by implicit feedback (such
as viewing behaviour), explicit feedback (such as ratings), or direct
user input (such as a semantic search query). This method is thus
versatile and applicable in a range of information retrieval
settings. Representing the different roles of a given entity as separate nodes
in a multilayer network provides unique opportunities for fine-grained
recommendations. For example, we can seed recommendations with Quentin
Tarantino as an actor or director depending on the preferences of the
user.

While intuitive, centrality measures based on random walks, such as
PageRank, suffer from localisation~\cite{Martin2014}, i.e. often most
of the mass of $\pi$ is associated with a small number of
highly-connected hubs. This behaviour is undesirable for thematic
recommendations because a large number of KG connections is a result
of rich knowledge about the node rather than an intrinsic measure of
quality (unlike in the original formulation of PageRank, in which
directed connections between websites can be seen as declarations of
interest or quality). To attenuate the effect of hubs on
recommendations, we can filter the ranked list of recommendations such
that
\begin{equation}
  z(v) = \left\{i \in z'(v) : \log_{10}\frac{\pi_i(v)}{\pi^*_i}
  \geq \theta\right\},\label{eq:filter}
\end{equation}
where $\pi_i(v)$ denotes the seeded PageRank score of node $i$, the
vector $\pi^*$ denotes the unseeded PageRank score 
(i.e. when $v_i = \frac{1}{N},\,\forall\, i$ ), and $\theta\in\R$ is a threshold. The
filtering in Eq.~\ref{eq:filter} serves to eliminate candidates that
are not sufficiently ``thematically'' related to the seeds; that is, their
PageRank score did not increase enough as a result of the
seeding (compared to the unseeded PageRank scores). For example, a
threshold $\theta=0$ ensures that the PageRank score of a candidate
does not decrease after seeding; $\theta=1$ keeps only nodes whose
score increased by at least an order of magnitude. A negative
$\theta$ tolerates some decrease in scores; this
can be useful in contexts where the presence of hubs or popular nodes
in the recommendations is desirable.


\section{Offline evaluation of thematic recommendations from user-item interactions}
\label{sec:eval}

\begin{figure}[tp]
  \centering
  \includegraphics[width=\textwidth]{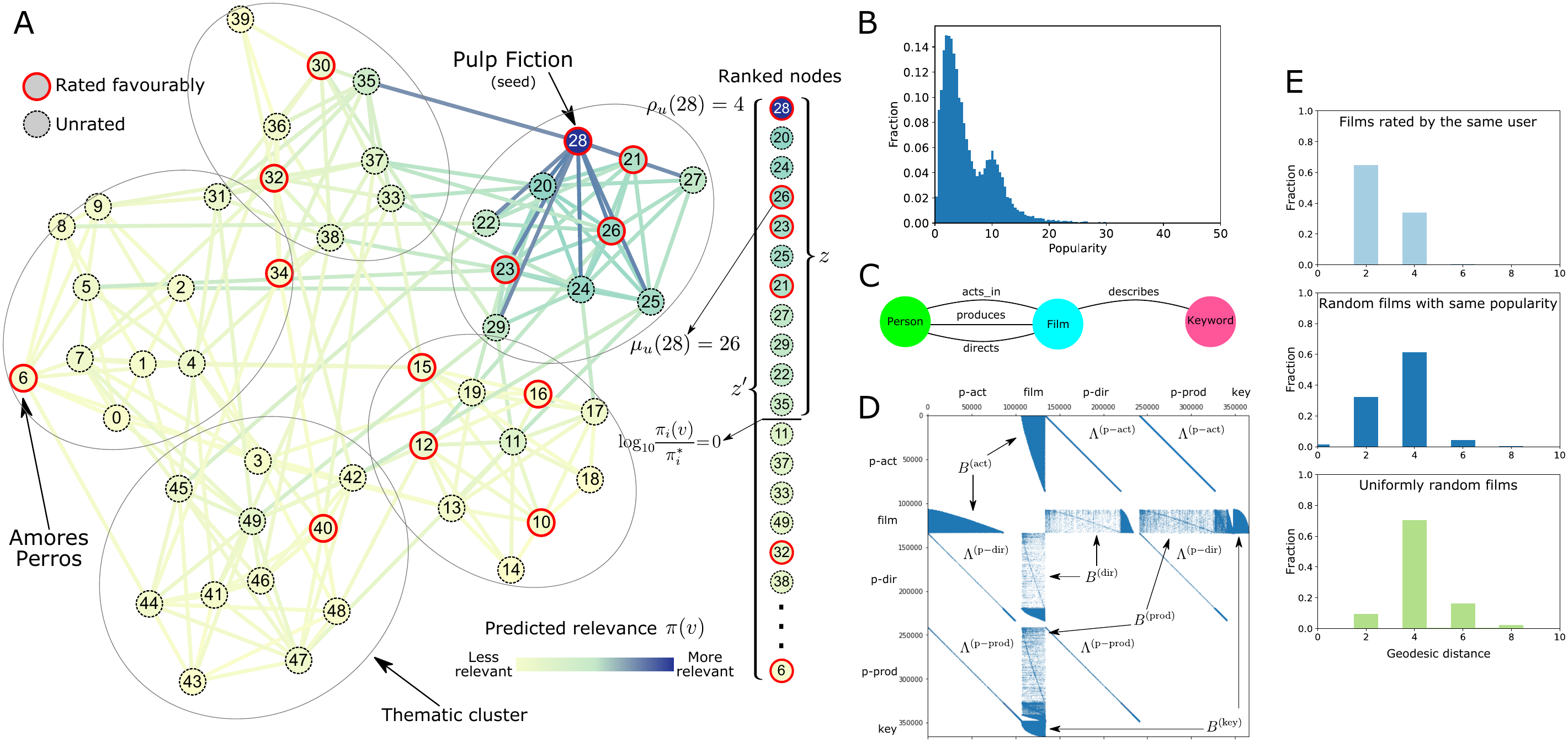}
  \caption{{\bf A}. An example network where connections between nodes denote a thematic relationship. 
    A user's favourably-rated
    items ($\mathcal{S}_u$) appear circled in red. The colour of each node
    represents its entry in $\pi(v)$. Seeding on Pulp Fiction (node 28) gives the
    ranking of the nodes $z'$ to the right, we also indicate the set
    $z$ of nodes whose PageRank increases as a result of the seeding
    (i.e. $\theta=0$ in Eq.~\ref{eq:filter}).  Given the seeding with
    Pulp Fiction, the highest-ranked item that was also rated positively by
    the user is node 26 at rank 4, i.e. $\mu_u(28)=26$  and $\rho_u(28)=4$.  {\bf B}.
    Distribution of popularity of films in TMDb.  {\bf C}. Schematic
    of the KG we construct from TMDb with three entity types and four
    connection types.
    {\bf D}. Sparsity pattern of the supra adjacency matrix of multilayer network from the TMDb KG.
    We indicate which blocks correspond to which layers and interlayer couplings.
    {\bf E}. Top: distribution of the shortest geodesic distance between films rated by 1,000 Movielens users (i.e. $\mathrm{dist}(s,s')$ where $s,s'\in\mathcal{S}_u$). Middle: shortest geodesic distance between
    films in 1,000 random sets of films with the same size and
    popularity distribution as the sets rated by the users in the top
    panel. Bottom: shortest geodesic distance between films in 1,000
    random sets of films with the same size as the user sets, where
    the films were chosen uniformly at random.}
  \label{fig:eval_fig}
\end{figure}

One of the best ways to evaluate the quality of a recommendation system is with an AB test, in which one cohort of users is exposed to recommendations from the system we wish to evaluate, and users from the control group receive recommendations from a system in place (or an alternative system). Although large-scale AB tests are an excellent way of evaluating the quality of recommendations, they are costly, time consuming, and risk sub-par user experiences for a proportion of users~\cite{Siroker2013}. As a result AB, tests are not practical for estimating good values of the unknown parameters of the network, such as the salience matrix $\alpha$ or teleportation probability $\rho$. In addition, AB testing may not be possible without access to a platform with enough users, which is also a barrier to researchers that wish to reproduce these results.

An alternative to AB tests lies in the judicious analysis of user-item interaction data. However, using such data is challenging because consumption  can be biased in favour of the recommendation system in place~\cite{Gilotte2018, Gruson2019}, governed  by personal user preference (as opposed to \emph{thematic} similarity), and driven by popularity~\cite{Beguerisse2010}. Nevertheless, thematic signals often do exist in user data (as we show in Sec. \ref{sec:thematic_signals_ind_user_data} for movie recommendation), although separating them from user preferences is non-trivial.

\subsection{Extracting thematic relevance from user preference data}
\label{sec:thematic_relevance}

We extract thematic information from user-item interaction data by relying on two main assumptions: 1) the KG is by construction ``thematically clustered'' (i.e, the KG has communities), and 2) users show affinity towards a number of thematic clusters (i.e., users have thematic preferences such as favourite genres, actors, directors and so on), which manifests itself as favourable ratings or increased consumption of items concentrated in these clusters.

Figure~\ref{fig:eval_fig}A illustrates these assumptions in a KG, where nodes are connected if they are thematically related (e.g. share an actor). The items rated favourably by the user (in red) are distributed across the KG, but are denser in the clusters that the user has most affinity to. In a traditional CF recommendation system, we would expect all the nodes in red to be near each other in a ranking. For example the 2000 film Amores Perros is often compared to Pulp Fiction (the seed)~\cite{Ebert2001}; it is likely that both films are rated similarly by several users, which a CF method would use as a signal to recommend them together. However, these two films are thematically distant and would be too far apart in the KG to be recommended together by our method.

Given a set $\mathcal{M}_{u}$ of items that user $u$ finds relevant (i.e. rated favourably or frequently consumed), we can use one item $m\in\mathcal{M}_{u}$ as a seed and record the rank of every other item $m'\neq m$, in the list of recommendations $z(v_m)$, where $v_m$ is the teleportation vector with most of its mass concentrated at item $m$, obtained by using Eq.~\eqref{eq:filter}. We denote these ranks as $r_{mm'}=\rank\left(m'|z\left(v_m\right)\right)$.
The higher the ranks $r_{mm'}$ are, the better recommendation list $z(v_m)$ is for this particular user. It would be tempting to use well-known scores such as the Mean Average Precision, or the Normalised Discounted Cumulative Gain (NDCG)~\cite{Manning2008} to quantify the quality of recommendations. These evaluations, however, do not take into account whether the observed user-item interactions were thematically driven, so they are less suitable for evaluating the quality of KG recommendations and penalise the normal behavior of our approach (i.e. not recommending thematically unrelated items). For instance, one obstacle are the many confounding factors that lead users to interact with a piece of content, such as user preference, or popularity. Therefore, we require a method that can pick up on the signal that we are most interested in, which in our case is thematic similarity.

There are two main challenges to isolating thematic signals from user-item interactions. First, it is difficult to identify the thematically relevant items-pairs, because identifying communities at the correct resolution in a multipartite, multilayer network, such as the ones we construct here, is not always practical. The second challenge is that it would be counterproductive to penalise thematically distant items that are correctly ranked lower in the recommendation list, even if they were favourably rated by the user. We assume that each item $m \in \mathcal{M}_u$ belongs to a thematic category (i.e. one of the KG's communities), whichever that may be, and that at least some thematic categories have multiple items rated by the user.
As we do not know which pairs in $z(v_m)$ are thematically relevant, we take an agnostic approach and consider only the highest ranked item favourably rated by the user. This item is likely to belong to the same thematic cluster as the seed; therefore, the position it appears in the ranked list can be used to evaluate the quality of our recommendation. In this way, we minimize the risk of penalising the algorithm for correctly down-ranking thematically distant items.

Specifically, our approach is to first select an item $m \in \mathcal{M}_{u}$ as a seed and generate a ranked list of recommendations $z(v_m)$, according to Eq.~\eqref{eq:filter}. The highest ranked item in $z(v_m)$ that is also relevant for our user $u$ is:
\begin{equation}
    \mu_{u}(m)=\argmin_{m'\in\mathcal{M}_u:m'\neq m}
    \rank\left(m'|z\left(v_m\right)\right),
  \label{eq:max_relev_item}
\end{equation}
and its rank is
\begin{equation}
    \rho_{u}(m)=\min_{m'\in\mathcal{M}_u:m'\neq m} \rank\left(m'|z\left(v_m\right)\right).
  \label{eq:max_relev_rank}
\end{equation}
Item $\mu_u(m)$ is the candidate most likely to be thematically related to the seed $m$ amongst the items $\mathcal{M}_u$ that are relevant to the user $u$. Focusing on $\mu_u(m)$ avoids penalising the method for correctly placing thematically distant items in $\mathcal{M}_u$ lower in the rankings. To evaluate the overall quality of recommendations for user $u$, we seed with every available item $m\in\mathcal{M}_u$ and compute an average score for each user which we call the \emph{Normalised Maximum Relevance Gain (NMRG)}:
\begin{equation}
  \mathrm{NMRG}_{u}=\frac{1}{\left|\mathcal{M}_{u}\right|}\sum_{m\in\mathcal{M}_{u}}\frac{1}{\varpi_{u}(m)}\frac{\tau_{u}\left(\mu_{u}\left(m\right)\right)}{\log_{2}\left(1+\rho_{u}\left(m\right)\right)},
  \label{eq:NMRG_user}
\end{equation}
where $\tau_{u}\left(\cdot\right)$ is the relevance of an item to user $u$ (e.g. a rating or affinity score), and 
\begin{equation}
    \varpi_{u}(m)=\max_{m'\in\mathcal{M}_{u}:m'\neq m}\tau_{u}\left(m'\right)
\end{equation}
is a normalisation constant representing the maximum possible value of the score,
corresponding to the case when the item with the highest relevance to the user appears in the first position of $z(v_m)$ (i.e. optimal outcome).

The NMRG considers only the position of the highest-ranked relevant item whereas the NDCG is a weighted average over the relevance of all items. Thus the NMRG is similar to the NDCG in the sense that it is proportional to the item's relevance to the user, inversely proportional to the log-rank, and normalised by the maximum possible score. The main difference is that the NMRG is no longer cumulative, but truncated to only one item: the \emph{highest-ranked relevant} one. Finally, note that a term in the sum of Eq.~\eqref{eq:NMRG_user} can only be zero if $\mathcal{M}_{u}\cap z\left(v_m\right)=\emptyset$, when an item $m$ has no thematically related counterparts in $\mathcal{M}_u$.

\section{Applications}
\label{sec:applications}

We test our thematic recommendation method in two different settings: thematic music recommendation, which we evaluate by monitoring user engagement in an AB test, and thematic movie recommendation, which we evaluate with the NMRG score. While the results of the AB test are not reproducible without access to a streaming service with a large user base, the film recommendations are performed on publicly available data, and can be reproduced by anyone.

\subsection{Thematic music recommendation: AB test}
\label{sec:ab-test}

We evaluate our thematic method for music recommendations. Specifically, we evaluate our KG-based recommendations in a new market for Spotify with two important characteristics that make it suitable for thematic recommendations: a) music consumption in this market is driven by thematic relationships among the entities in a KG which include people, music and movies, and b) the market comes with a new catalogue, so there is little historical user consumption data (i.e. cold market and cold catalogue). We performed an AB test where we compared the performance of playlists obtained with our KG-based method (as described in Sec.~\ref{sec:network_mods} and~\ref{sec:random}), and the CF system that was in production at the time. Our hypothesis is that we can better capture the dynamics of consumption in this market/catalogue with our thematic approach than with the CF system.
We produce song-song recommendations by creating a playlist from each seed (i.e., a track) in the following way: we first obtain 300 candidates using equation~\eqref{eq:filter}. Then, we reduce the candidate list to a 40-song playlist by re-ranking candidates to enforce artist/album separation rules, and ensure mood consistency using audio attributes. We set the value of the salience parameters by working closely with an editorial team with expertise in this market and deep knowledge of the catalogue.
\begin{figure}[tp]
  \centering
  \includegraphics[width=0.5\textwidth]{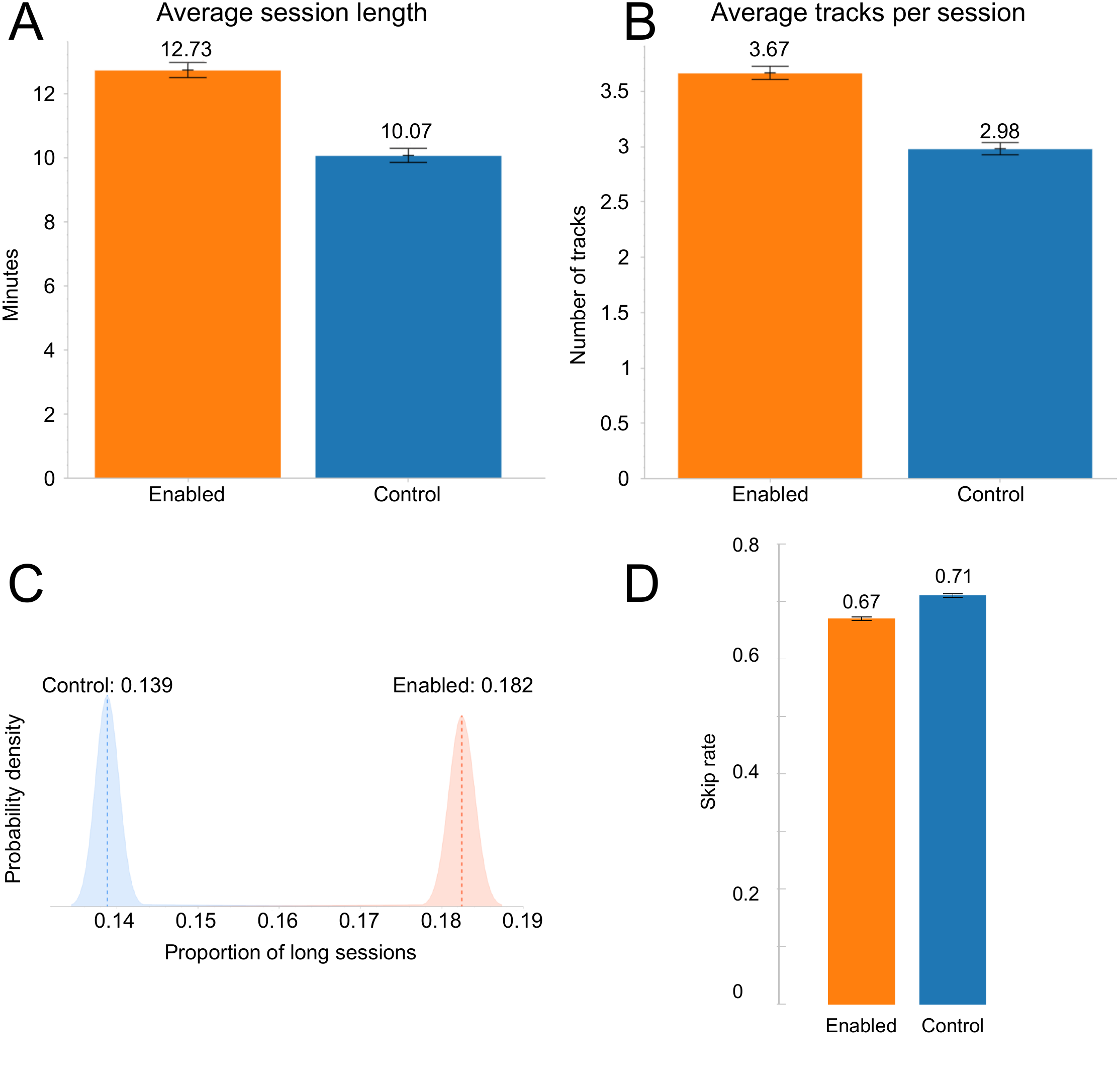}
  \caption{AB test results. {\bf A}: average session length in minutes. {\bf B}: average number of tracks per session. {\bf C}: proportion of sessions of five tracks or more. {\bf D}: skip rate.}
  \label{fig:AB_test}
\end{figure}
For the test, we assembled a cohort of 100,000 users, 50\% of which received our KG-based recommendations, and a control group of 50\% received the CF recommendations. After two weeks, the group exposed to KG reported (see Fig.~\ref{fig:AB_test}):
\begin{itemize}
    \item 26\% increase in consumption time compared to control,
    \item 31\% increase in users with sessions of 5 tracks or more,
    \item 6\% decrease in skip rate, and
    \item 5\% increase in save rate.
\end{itemize}
We replicated these results in two further AB tests with the same sample size, which gave us the confidence to roll out the system to all users, and this is currently the recommendation system in production in this market.

\subsection{Thematic movie recommendations: offline evaluation}
\label{sec:tmdb_application}

Next, we test our thematic KG recommendation method for movie recommendations. In this application, we combine data from two different sources: the Movie Database (TMDb)~\cite{tmdb_website} metadata for the construction of a KG, and the MovieLens 20M dataset~\cite{Harper2015} as a source of explicit user feedback.

We obtained metadata from TMDb's API~\cite{tmdb_website} for all movies in the MovieLens dataset with a valid TMDb identifier. These data capture extensive information on how different types of entities (movies, people, and companies) are interrelated. For example, there are over 800 job titles available, including categories such as cast member, director, producer, and even animal wrangler. The data also include the order in which actors appear in a movie's credits, which we use as a proxy measure of importance of an acting credit. Additionally, the dataset contains 17k unique keywords of which 10k are associated with more than one movie. Finally, each movie in the TMDb data is associated with a popularity score, based on engagement metrics, such as votes, views, number of users who marked it as favourite, users who added it to their ``watchlist'', and release date~\cite{tmdb_website}. Figure~\ref{fig:eval_fig}B shows the distribution of these popularity scores, which we incorporate into the weight of the KG connections elevated to an exponent which is also a parameter. For full details on the graph construction we refer to Appendix~\ref{appendix:KG_contruction}.

The MovieLens 20M dataset~\cite{Harper2015} comprises 20 million user ratings of 27k movies by 138k users; each rating has an integer value between 1 and 5.  The data contain TMDb identifiers for most movies and 12 million associations of 15k movies with 1,100 tags, capturing properties of movies such as sci-fi, comedy, action, surreal, funny, classic, romance~\cite{Vig2012}. See Appendix~\ref{appendix:movielens} for the details on the data preprocessing and train/test set splits.

\subsubsection{Thematic signals in user data} \label{sec:thematic_signals_ind_user_data}

In Sec.~\ref{sec:thematic_relevance}, we hypothesised that users display thematic preferences (such as favourite directors, actors, genres and so on); we can test this using user-item ratings in combination with our KG. If our hypothesis is correct, we expect that the path in the KG between movies rated favourably by the same user to be shorter than between two movies chosen at random. To test this hypothesis, we sample the sets of ratings ($\mathcal{M}_u$) of 1,000 users, and measure the shortest path in the KG between all pairs $(m, m') \in \mathcal{M}_u$. We compare these to the paths between sets of movies with the same popularity as in $\mathcal{M}_u$, and between movies chosen uniformly at random from the 27k movies in the dataset.
Figure~\ref{fig:eval_fig}E shows the distribution of shortest paths between movies in the three sets of ratings from $1,000$ users each. The distances between movies favourably rated by users are shorter than between random movies, which confirms that users do tend to interact with movies that are \emph{thematically similar} to each other. This result shows that we can find thematic signals in user-item interaction data.

\subsubsection{Parameter estimation} \label{sec:kg_parameter_estimation}

Our multilayer models of KGs have several free parameters including saliences and the PageRank teleportation probability, whose value may have a large impact on the quality of recommendations, and are often unknown a priori. We use the NMRG evaluation framework from Sec.~\ref{sec:eval} along with random search~\cite{bergstra2012random} to find parameter values that optimise the thematic relevance of recommendations, as measured by the NMRG score. The random walk model form the TMBDb data KG (Fig.~\ref{fig:eval_fig}C and~D) has 14 free parameters (13 saliences and the PageRank teleportation). We sample each column of the salience matrix in Eq.~\eqref{eq:salience-matrix} from a Dirichlet distribution with dimensions equal to the number of corresponding non-zero blocks. For example, the dimensionality of the Dirichlet random variable for the first column in Eq.~\eqref{eq:salience-matrix} is three. We draw each column from a flat Dirichlet distribution and the teleportation parameter from the uniform $\mathcal{U}(0,1)$ distribution.
For each randomly sampled parameter configuration, we compute recommendations for every one of the $103.9k$ users in the training set, and average the NMRG across all users at 1, 10, and 20 recommendations.
We then fine tune the teleportation probability by performing a sweep using the best configuration from the Dirichlet sampling.
See Appendix~\ref{appendix:random_search} for more details on the parameter search and a discussion of our findings. 

\subsubsection{Comparison with baselines and alternative recommendation systems} 
\label{sec:hypotheses_testing}

We compare the quality of our recommendations against four baselines and three alternative recommendation methods:
\begin{enumerate}[a)]
    \item \emph{Popularity baseline}: the recommendations are the same for every seed $m$. The ranking consists of the $K\in\{1, 10, 20\}$ most popular items in the catalogue, excluding the seed. This baseline represents the hypothesis that user engagement is 100\% driven by popularity.
    \item \emph{Random seed baseline}: we replace each seed $m$ with a random item of \emph{similar popularity}\footnote{\label{ref:sim_pop} Items of similar popularity are randomly selected from the 25 items with the closest popularity score to $m$.}. This baseline represents the  hypothesis that user engagement is driven by the popularity of the seed rather than the seed itself.
    \item \emph{Random item baseline}: we retain the original seed $m$ but replace each recommendation in $z(v_m)$ by a random item of \emph{similar popularity}. This baseline represents the hypothesis that user engagement is driven by the popularity of the recommended items.
    \item \emph{Unseeded baseline}: we rank items according to their unseeded PageRank (i.e. uniform teleportation). This baseline represents the hypothesis that the node centrality in the KG drives user engagement.
    \item \emph{TMDb keywords}: we rank items based on the similarity of their associated TMDb keywords, as measured by the Dice score~\cite{Carass2020}.
    \item \emph{Movielens tag genome}: we use the graded associations between films and keywords~\cite{Harper2015} provided by the Movielens dataset to rank candidates by the Euclidean distance of their association vector and the seed's.
    \item \emph{SVD}: we use the open-source library \texttt{surprise}~\cite{Hug2017} to obtain a low-rank approximation of the ratings matrix using Singular Value Decomposition (SVD). In this approach, we rank items based on their dot-product similarity to the seed.
    \item \textit{TMDb CF}: we use the collaborative filtering recommendation currently in use on the TMDb website~ \cite{tmdb_website}.
\end{enumerate}

\begin{table}
  \begin{tabular}{ | l | c | c | c |}
    \hline
    \textbf{Method} & \textbf{NMRG@1} & \textbf{NMRG@10} & \textbf{NMRG@20} \\
    \hline
    \small Thematic KG recs & $\boldsymbol{14.83 (\pm0.16)}$ & $\boldsymbol{30.70 (\pm0.25)}$ & $\boldsymbol{35.07 (\pm0.24)}$ \\
    \hline
    \small Popularity baseline & $0.53 (\pm0.09)$ & $19.77 (\pm0.23)$ & $22.02 (\pm0.21)$ \\
    \small Random seed baseline & $ 4.91 (\pm0.07)$ & $ 16.28 (\pm0.17)$ & $ 20.86 (\pm0.18)$ \\
    \small Random item baseline & $7.36 (\pm0.10)$ & $21.95 (\pm0.21)$ & $27.28 (\pm0.21)$ \\
    \small Unseeded PageRank  baseline & $0.70 (\pm0.11)$ & $23.60 (\pm0.25)$ & $26.95 (\pm0.21)$ \\
    \hline
    \small TMDb keywords & $4.44 (\pm0.08)$ & $7.34 (\pm0.10)$ & $8.57 (\pm0.10)$ \\
    \small Movielens tag genome & $12.61 (\pm0.13)$ & $22.84 (\pm0.18)$ & $25.02 (\pm0.18)$ \\
    \small SVD & $14.46 (\pm0.14)$ & $\boldsymbol{30.70 (\pm0.20)}$ & $33.22 (\pm0.19)$ \\
    \small TMDb CF & $14.36 (\pm0.15)$ & $28.68 (\pm0.20)$ & $31.14 (\pm0.20)$ \\
    \hline
  \end{tabular}
  \caption{Offline evaluation NMRG scores at 1, 10, and 20 items, for all methods and baselines.}
  \label{table:nmrg_results}
\end{table}

We report NMRG scores at 1, 10, and 20 recommendations calculated on the test set only. We compare all methods using the same data, except the MovieLens tag genome and SVD, where only a subset of movies is available. We remove ratings that are not available for evaluation to avoid penalising these methods, even though this may lead to results slightly skewed in their favour. Table~\ref{table:nmrg_results} contains the average NMRG score for each method with 99\% confidence intervals.
The performance of the baselines highlights the importance of the information encoded in the KG (in line with our findings in Fig.~\ref{fig:eval_fig}E); the best performing baselines are the ones that exploit the structure of the KG.
Among the alternative methods, recommending based on TMDb keywords has the poorest performance, followed by the Movielens tag genome. The TMDb CF recommendation performs better than all baselines and is only outperformed by the SVD approach, which is tuned on the Movielens ratings. Our thematic recommendations performs as good as the SVD method or better in some cases.
However, it is important to note that we do not claim superior performance to CF methods on \emph{personalisation}, and the NMRG is specifically tailored to highlight the \emph{thematic} aspects of the KG recommendations.

\section{Discussion}
\label{sec:discussion}

In this work, we introduce a method for thematic item-item recommendations using knowledge graphs. We represent a KG as a multilayer network whose layers correspond to different connection types, and interlayer couplings connect nodes that represent the same entity across layers.  By incorporating parameters that encode the salience of each of the connections, we are able to compare different types of connections on the same footing, allowing us to unify different connection types under the same analysis framework.  Representing the KG as a multilayer network enables us to draw from a wealth of network theoretic techniques, such as random walks. We use personalised PageRank to produce recommendations that can be fine-tuned by calibrating the salience of the connections or the teleportation probability, which controls how exploratory recommendations are.

We evaluated our method for music recommendations in a specific market using an AB test. Our results show that our thematic recommendations perform significantly better than the CF in place in a range of engagement metrics. With this strong performance 
we rolled out the system for all users in this market.
We also evaluated our framework for movie recommendations on publicly-available data. We built a KG from TMDb data and tested the performance of our method using the NMRG metric we introduce in this paper on the MovieLens 20M dataset. The evaluation framework also allows us to calibrate the parameters in the model to optimise the NMRG score and discriminate among competing models. We perform random search and parameter sweeps to identify parameter sets with good performance, and find that our method outperforms both thematic and collaborative filtering methods.

An important challenge arises during the evaluation of thematic recommendations: it is difficult to disentangle the effect of item popularity from user-item interactions. Our method performs best after incorporating the TMDb item popularity score in the connection weights. The more we accentuate the popularity, the better the model performs. However, recommending items purely based on popularity as a baseline does not perform well. This is evidence that the combination of the graph structure and popularity is required for high user satisfaction, which is also consistent with our
experience in commercial applications.

Our framework is flexible and allows more complex random walk dynamics to be easily incorporated. For example, we can encode higher-order Markov chains~\cite{Rosvall2014, Lambiotte2019} at the layer level by constructing more granular roles (e.g. Tarantino acting in the 90s, Tarantino acting in the 00s and so on) and manipulating the values in the salience matrix $\alpha$, although this comes at the cost of a larger system. We used the steady state of a diffusion process to recommend items. However, transients of diffusion in discrete or continuous time offer attractive alternatives. Transients do not require irreducibility of the rate matrix, and extremely fast numerical solvers are readily available. The duration of the diffusion can be optimised globally or, more interestingly, be personalised to users, with longer durations for more adventurous users. See Ref.~\cite{Nikolakopoulos2019} for a recent example in discrete time.

Our framework is currently limited by the fact that it is challenging to evaluate purely thematic recommendations (i.e. without popularity in the model) due to the lack of datasets where the influence of popularity is absent. Therefore, an important task that remains is the creation of purely thematic datasets and developing evaluation methods to disentangle thematic relatedness and popularity. Finally, it is possible to integrate CF approaches into this framework by adding an item-user layer to the model of the KG. This is a promising direction of research, as it combines thematic and collaborative recommendations in a principled, transparent framework that can be understood, manipulated and optimised.

\section*{Acknowledgments}
We thank Jonathan Berschadsky, Maria Dominguez, Katarzyna Drzyzga, Shahar Elisha, Martin Gould, Johnny Hunter, Nachiket Londhe, Laurence Pascall, Jyotsna Venkataramanan, and Linden Vongsathorn for their help and support as well as their useful feedback and comments. We are also thankful to the developers of TMDb for kindly providing us with support and access to their API.

\begin{appendices}

\section{Movie data and preprocessing} \label{appendix:data}

\subsection{Construction of the knowledge graph} \label{appendix:KG_contruction}

The TMDb Knowledge Graph includes $P=107,143$ people, $F=26,698$ movies, and $K=17,592$ keywords. We construct the multilayer model of the KG with three roles for people entities: actor (\role{act}), director (\role{dir}), and producer (\role{prod}). We also consider a role (\role{desc}) for keywords describing each movie. We represent movies as a single \role{movie} role instead of modelling movies in four distinct passive roles (i.e. \role{m-dir}, \role{m-prod}, \role{m-act}, and \role{m-key}). As discussed in Sec.~\ref{sec:network_mods}, merging roles reduces the size of the supra adjacency matrix (in this case by 80k rows and columns), which helps speed up computations. Figure~\ref{fig:eval_fig}D shows the sparsity profile and block structure of the supra-adjacency matrix, comprising four distinct layers. We experimented with several versions of the KG with different types of connections\footnote{Including animal wrangler, of course.}. For brevity, we only present the configuration that performed best here.

We incorporate the popularity of the movies by letting the weight of connections to each movie $m$ to be proportional to $p_m^\gamma$, where $\gamma\in\R$ is an exponent controlling the importance of popularity. For instance, $\gamma>1$ accentuates differences in popularity, $\gamma=0$ discards them, and $\gamma<0$ reverses their effect. We set the weight of \role{directs}, \role{produces} and \role{describes} connections to $p_m^\gamma$. The weights of \role{acts\_in} connections are set to $\frac{p_m^\gamma}{\log_2(c_i + 1)}$, where $c_i$ is the position of actor $i$ in the credits, to account for both popularity and credit order. We observed that higher values of the popularity exponent $\gamma$ result in increased performance, and fixed this parameter to a large value ($\gamma=30$). This highlights the overwhelming importance of popularity, and in effect sparsifies the network by connecting each node to its' most popular associated movie only.

\subsection{MovieLens 20M dataset} \label{appendix:movielens}

To obtain the data we use in our experiments, we apply heuristic filtering steps to the user ratings in the order shown in Table.~\ref{table:filtering}. The total number of users is not affected by the filtering. More specifically, we retain only favourable user ratings (above the user's median rating) and remove the rest to avoid rewarding methods that recommend items the users may not have liked. We also remove movies with 2 ratings or fewer, because we expect those ratings to be noisy. Finally, we keep only the 250 maximum ratings per user, so that we do not bias our metrics in favour of power users.

We split the user set uniformly at random into a train (75\%) and test set (25\%). We use the train set for random parameter search and training. We report results on the test set only for all methods. The resulting train set  comprises $103.9k$ users, $17.6k$ movies and $7.9M$ ratings, while the test set consists of $34.6k$ users, $15.1k$ movies and $2.6M$ ratings.

\begin{table}
    \centering
    \begin{tabular}{| l | l || c | c |}
        \hline
        \textbf{Step} & \textbf{Description} & \textbf{\#ratings} & \textbf{\#movies} \\ \hline \hline
        \#1 & Remove ratings without TMDb id & 19,987,681 & 26,483 \\ \hline
        \#2 & Remove duplicates (keeping latest) & 19,987,649 & 26,483 \\ \hline
        \#3 & Remove movies without metadata & 19,950,334 & 26,185 \\ \hline
        \#4 & Remove ratings below user median & 13,032,003 & 23,431 \\ \hline
        \#5 & Remove movies with 2 or fewer ratings  & 13,028,453 & 19,881 \\ \hline
        \#6 & Keep only top 250 ratings per user & 10,563,717 & 18,319 \\ \hline
    \end{tabular}
    \caption{Filtering steps applied to the dataset, with their corresponding impact on the number of ratings and number of movies.}
    \label{table:filtering}
\end{table}

\begin{table}[tp]
 \centering
  \begin{tabular}{ | l || c | c | c | c | c |}
    \hline
     & \textbf{Actor} & \textbf{Movie} & \textbf{Keyword} & \textbf{Producer} & \textbf{Director}\\ \hline \hline
    \textbf{Actor} & - & $0.798$ & - & $0.643$ & $0.149$ \\\hline
    \textbf{Movie} & $0.3585$ & - & $1$ & $0.057$ & $0.424$ \\ \hline
    \textbf{Keyword} & - & $0.012$ & - & - & - \\ \hline
    \textbf{Producer} & $0.550$ & $0.064$ & - & - & $0.427$ \\ \hline
    \textbf{Director} & $0.091$ & $0.126$ & - & $0.299$ & - \\ \hline
  \end{tabular}
  \caption{Best salience values discovered using random search.}
  \label{table:random_search_vals}
\end{table}

\begin{figure}[tp]
  \centering
  \includegraphics[width=1.0\textwidth]{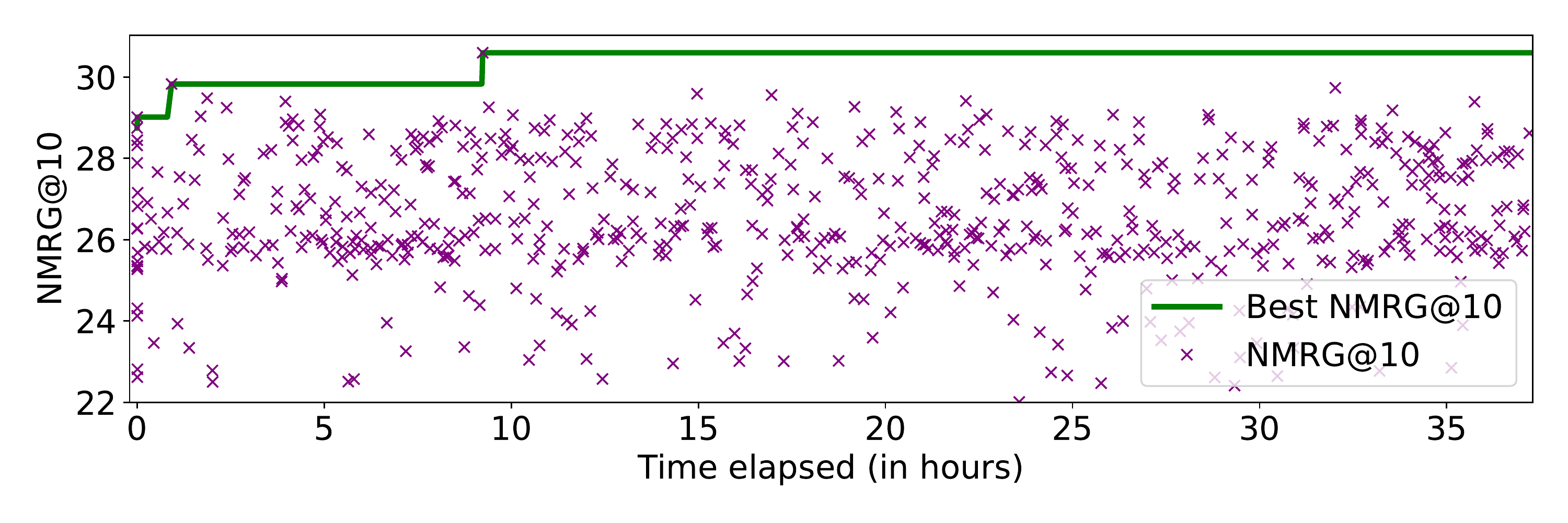}
  \caption{Random search NMRG@10 and best NMRG@10 as the experiment evolves.}
  \label{fig:saturation}
\end{figure}

\section{Random parameter search} \label{appendix:random_search}

We use the evaluation framework introduced in Sec.~\ref{sec:eval} along with random search~\cite{bergstra2012random} to find good values of the 14 free parameters in our model.
The teleportation parameter $\rho$ in Eq.~\ref{eq:teleportation-evolution}, takes positive values between $(0, 1)$, so we sample it from uniform distribution $\rho\sim\mathcal{U}(0, 1)$. 
We sample the values of each column in Eq.~\ref{eq:salience-matrix} from a Dirichlet distribution with unit concentration parameter $\boldsymbol{\alpha_i}\sim Dir\left(\boldsymbol{1}\right)$, where $i$ is the column index.

Evaluating on the full train set requires a considerable amount of computation, despite the fact that the process can easily be performed in parallel. To optimise computations, we use a two-stage approach: we first compute ranked lists $z(v_m)$ by seeding at each one of the 17.6k movies in the training set, then we use these lists to evaluate the NMRG for each user following Eq.~\eqref{eq:NMRG_user}. Such pre-computing speeds up our calculations significantly. We present the best salience parameters as identified by the random search in Table~\ref{table:random_search_vals}.

The results of our random parameter search were obtained after evaluating 655 random parameter configurations. For this computation we employed 5 machines with 96 vCPUs each; the total run time was approximately 37 hours. In Fig.~\ref{fig:saturation} we present the NMRG@10 and best NMRG@10 as the experiment evolves.

\begin{figure}[tp]
  \centering
  \includegraphics[width=0.5\textwidth]{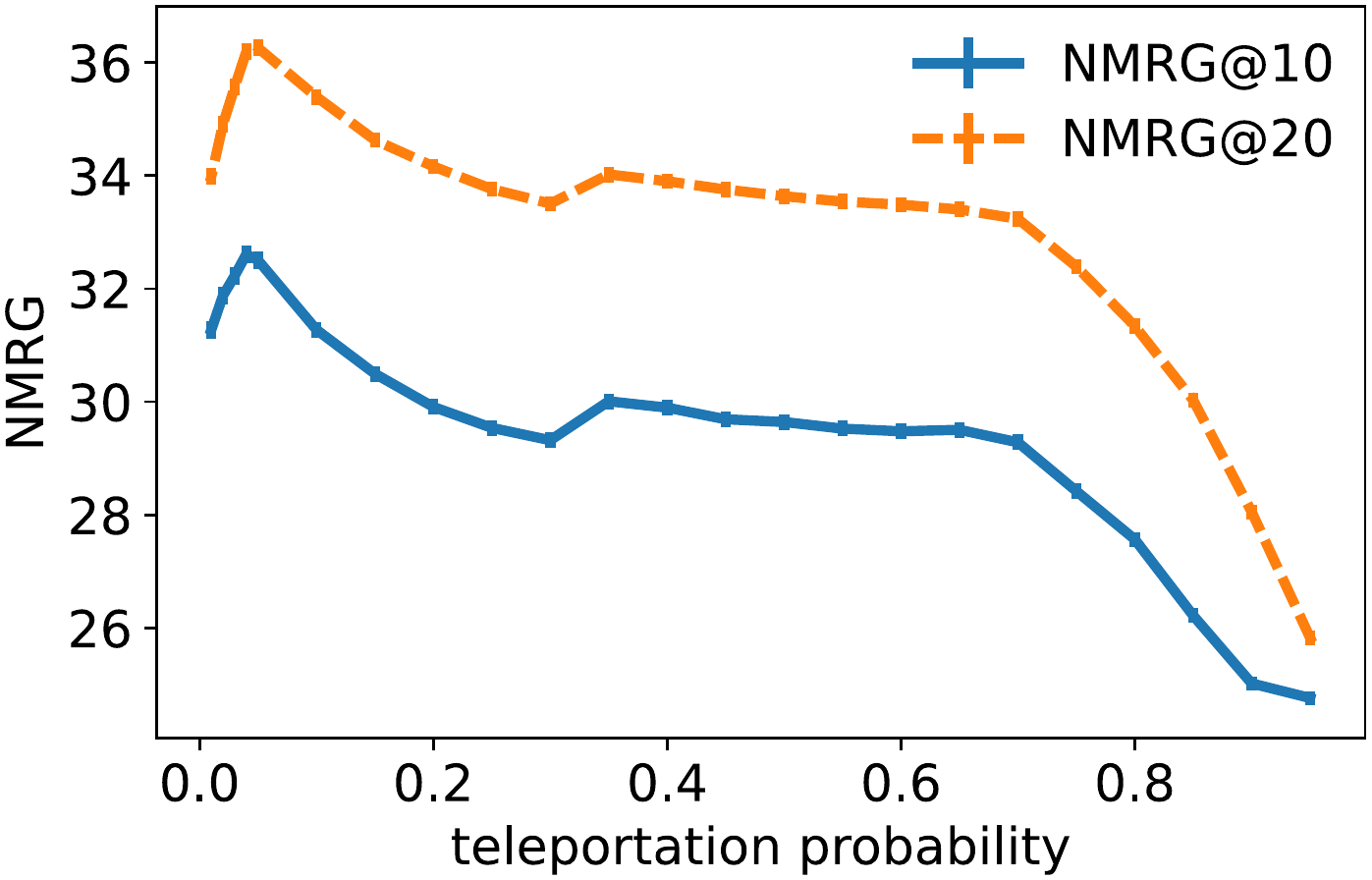}
  \caption{Teleportation parameter ($\rho$) sweep.}
  \label{fig:teleport_sweep}
\end{figure}

Figure~\ref{fig:teleport_sweep} shows a sweep of the teleportation probability $\rho$. The optimal value of the NMRG is achieved when $\rho\approx 0.12$.

\end{appendices}

\bibliographystyle{siamplain}

\end{document}